# Department of Botany

## Netrakona Govt. College, Netrakona

RESEARCH REPORT ON

**The use of Ethnomedicinal plants in Indigenous Health Care Practice of the Hajong Tribe community in Durgapur, Bangladesh**

Research Report Presented to the Department of Botany .

## Supervised by

**Prof. Mohammad Manirul Islam**

Head

Department of Botany

Netrakona Govt. College, Netrakona

### Submitted by

### Ashik Saha

REG. No: 18330109392

Session: 2018 - 2019

### Date of submission:

## Table of Contents

# Acknowledgments

*"This research is dedicated to my parents, family members, teachers and my friends and all the people who are involved to make it successful and always inspired me in every step to accomplish this study".*

At first, I would like to pass my gratitude and thanks to my honorable Supervisor **Prof. Mohammad Manirul Islam** for his valuable suggestions, ideas and guidance in every step of my work to accomplish the study. I am very grateful to him for giving me the opportunity to work with him. Without his continuous support and guidance, it would have been very difficult for me to complete this thesis. Our respected faculty members deserve my heartfelt thanks for their guidance to make proper documentation of my paper. I must mention my honorable Principal, **Prof. Nurul Baset** for sharing his wisdom before and during this study. Finally, I would like to express my gratitude to Netrakona Govt. College for providing me an excellent environment for research.

## ACCEPTING GRATEFULNESS

I am grateful to the Most Gracious Almighty. Then Principal of Netrokona Government College **Professor Md Nurul Baset sir**. **Professor Md. Monirul Islam**, Divisional Head of Department of Botany, expressed his gratitude to Sir, because without the help of Sir, my research report could not be prepared.

I thank my supervisor teacher Sir, from the heart of my heart. Thank you very much for making the research report easy. It is possible to make the report easy for its constant cooperation.

I also thank The Hajong people, who were very cooperative to us to share their ethno-medicinal knowledge with us. I are very grateful to **Salma Khatun** and **Manisha Hajong** for their help in our exploration of the plants used by the tribal peoples.

# Abstract

**Background**: The Garo Hills have always been fascination to the naked human eyes. The hills are the shelter of the earliest human habitation of Bangladesh. It is a place of ancient cultures and many botanical wonders. It is situated in the most northern part of Durgapur sub-district having border with Meghalaya of India. Durgapur is rich with ethnic diversity with Hajong and Garo as the major ethnic groups along with some Bengali settlers from the common population. Present survey was undertaken to compile the medicinal plant usage among the Hajong Tribe of Durgapur.

**Methods**: The ethnomedicinal data was collected through open and focused group discussions, personal interviews and from The Tribal Medicine Practitioners (TMP) using a semi-structured questionnaire. Before collecting data, I took consent from all the tribal people. They all were very cooperative and supportive. After that I put the data in a structured format and analyzed the information. I have found various plants and animal parts that used in some common diseases as local ailment.

**Results**: In our studied area I have found a total of 39 plants from 31 families. I have documented and identified all of the plants. Gastrointestinal disorders represented the major ailment category with use of 39 plant species followed by Respiratory tract infection and Intestinal worm problem.

**Conclusion**: Present investigation on Hajong tribe was very successful. I have found a rich traditional practice and use of enormous ethnomedicinal plants. The studied region was very profound and resourceful. The compilation of the ethnobotanical knowledge can help other researchers to identify the use of various medicinal plants that have a long history.

## BACKGROUND

Man has been using the plants as medicine from the earlier times and present years, that interest the people to investigate plants for use of herbal medicine, pharmaceuticals, flavorings, perfumes, cosmetics and another natural product (Iqbal, 1993; Walter, 2001 Rao and Arora, 2004). Tribal people are the people who are close to the nature most. Foods, fodders, medicine and other forest products have made a traditional economy of the tribal communities of the world (Miah and Chowdhury, 2003). Prakash (1999) estimated that by consulting indigenous peoples, bio-prospectors could increase the success ratio in trials from one in 10,000 samples to one to two. He also documents that traditional knowledge increases the efficiency of screening plants for medicinal properties by more than 400 percent. If such inputs of indigenous knowledge are not involved throughout the world, many valuable medical products used extensively today, would not exist in the future. Through evolution, plants have developed large numbers of chemical substances to defend themselves against insect pests, and fungal and other pathogenic diseases

The Garo Hills in Durgapur sub-district is one of the most remote areas of the northern part of Bangladesh. Ethnic groups like Garos and Hajongs reside in this area from ancient times along with the Bengali common people who also had settle them in this region hundreds of years ago. The Hajongs are a small indigenous tribal community residing in the hills of this forested area. They mostly live in the north central districts of Bangladesh like Mymensingh, Netrakona, Sherpur and Jamalpur. They are known to arrive here from Tibet through Assam, India. They are mostly Hindus by religion. According to their neighboring tribe, The Garos, 'Ha' means soil and 'Jong' means insect. Since the Hajongs are mainly agriculturists, The name Hajong is derived from the two words, which cumulatively means people working with the soil. According to the Hajongs, the word 'Hajong' means dressing up, particularly being attired in battlefield dress.

The Hajongs are the ancient among the ethnic minorities of the North-Eastern region of Bangladesh who represent different types of socio-political organization compared to the other ethnic group (Nasrin and Khalifa, 2004). The total population of the Hajong in Bangladesh is 11,477 (Anon, 2003). Although the Hajongs, in recent years, are fast losing their cultural identity because of association with the mainstream Bengali-speaking population, they still cling to some of their traditional rituals and customs, including their traditional medicinal practices. This medicinal practice, is to our knowledge, not previously been documented. Indigenous communities, through long association with plants acquired by living in forested areas, possess quite extensive knowledge about the medicinal properties of plants and other denizens of the forest. Documentation of this knowledge is important because many useful modern drugs like atropine, reserpine, strychnine, quinine and artemisinin, to name only a few have come from close observations of indigenous community practices (Balick and Cox, 1996; Cotton, 1996; Gilani and Rahman, 2005). Such information from ethnic groups or indigenous traditional medicine has also played a vital role in the discovery of novel chemotherapeutic agents from plants (Katewa et al., 2004). I had been conducting extensive ethnomedicinal surveys among the traditional practitioners of folk medicine (Kavirajes) in Bangladesh as well as tribal medicinal practitioners (TMPs) of various tribes for the last few years to document this ethnomedicinal knowledge (Nawaz et al., 2009; Rahmatullah et al., 2009a-c; Chowdhury et al., 2010; Hasan et al., 2010; Hossan et al., 2010; Mollik et al., 2010a,b; Rahmatullah et al., 2010a-g; Akber et al., 2011; Biswas et al., 2011a-c; Haque et al., 2011; Islam et al., 2011; Jahan et al., 2011; Rahmatullah et al., 2011a,b; Sarker et al., 2011; Shaheen et al., 2011; Das et al., 2012; Rahmatullah et al., 2012a-d). The objective of the present study was to conduct an ethnomedicinal survey among a Hajong community located in Baromari village of Durgapur Upazila (sub-district) in Netrakona district of Bangladesh. In our previous ethnomedicinal surveys, I have found wide variations in the selection of medicinal plants by Kavirajes living even in adjacent villages, or TMPs of the same tribe but living in different locations. As such, I feel that it is important to survey all TMPs or Kavirajes to get a comprehensive view of medicinal plants used in folk and tribal medicines of Bangladesh.

## METHODOLOGY

### Study area

The study was conducted at Gopalpur and Baromari union (a remote administrative unit consist of a number of villages) of Durgapur Upazilla (Sub-district) of Netrokona district, Bangladesh. Its coordination is 25.1059° N, 90.6714° E . It's a bit remote from the center of Netrokona. According to 2011 Bangladesh census, Durgapur had a population of 224,873. Males constituted 49.67% of the population and females 50.33%. Muslims formed 90.10% of the population, Hindus 5.87%, Christians 3.89%, and others 0.14%. Durgapur had a literacy rate of 39.52% for the population 7 years and above. At the 1991 Bangladesh census, Durgapur had a population of 169,135, of whom 83,795 were aged 18 or older. Males constituted 50.49% of the population, and females 49.51%. Durgapur had an average literacy rate of 23% (7+ years), compared to the national average of 32.4%.

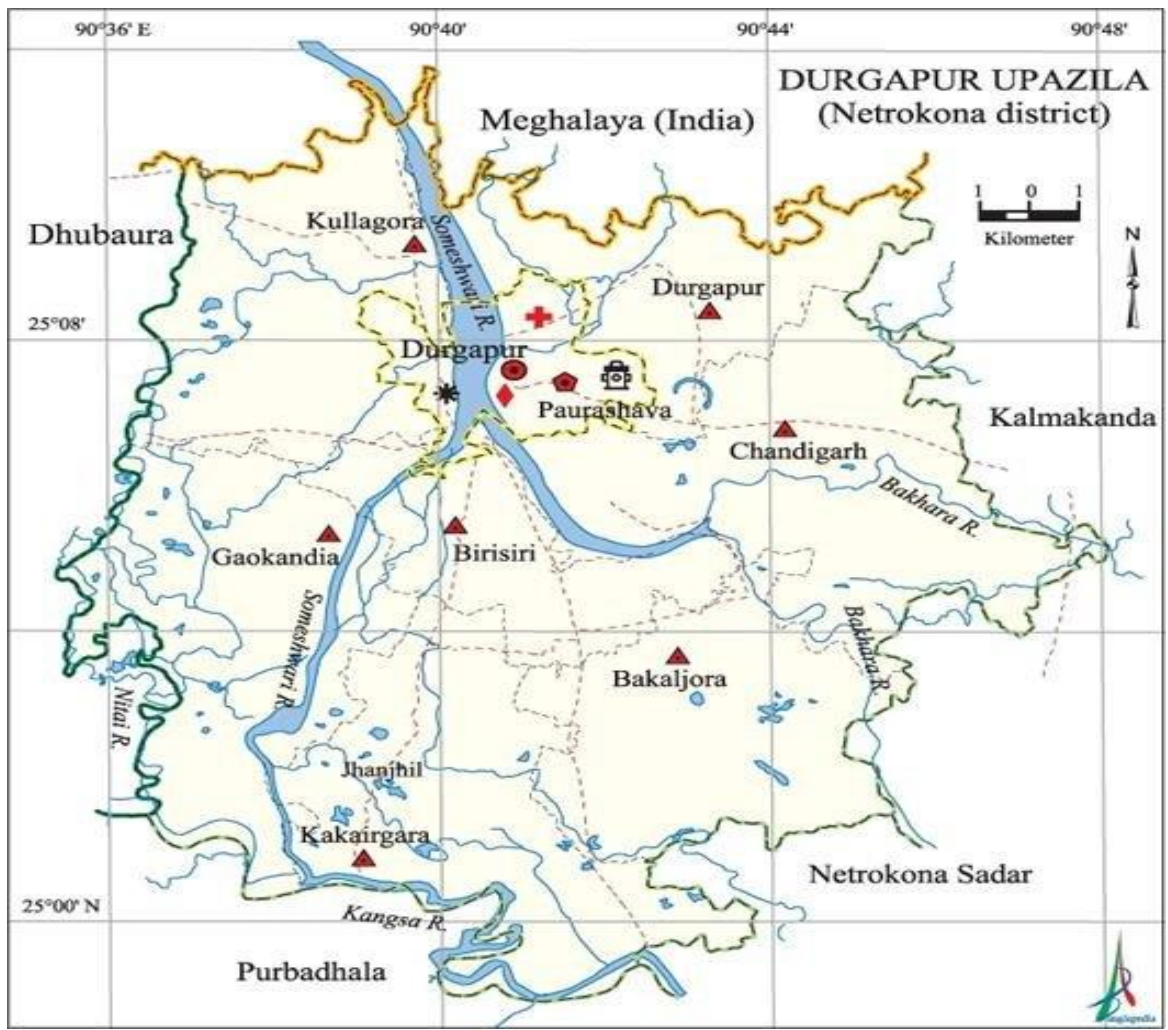

*Fig. 1: Map of Study Area*

**Method**

The Hajongs are one of the most ancient among the tribes of Northern-Eastern region of Bangladesh. They mainly found in these regions of the country. Netrokona District was selected by our Mentor, as it is not very far from our living place an not been discovered very carefully. Hajong tribe has a profound knowledge of herbal medicine that could help us in the research of our modern medicine development.

I had a semi- structured questionnaire to have an ease on our data collection. I had a local guide to take us to the places where most of the hajong tribes will be. I asked all the people with consent if the want to share their ethnomedicinal knowledge with us. They were interviewed using a semi-structured questionnaire to ascertain the plants they use, parts they use, for what diseases, sources they prefer, the reason for cultivating any plant and the eagerness of the younger generation in this regard. The plant species used for medicine were firstly identified by local names. The scientific names were obtained by consulting the literature (BARC, 1972-1992; Chopra et al., 1992; Chevallier, 1996; Das and Alam, 2001). A final list of the species used for medicinal purposes was prepared based on the study by Dey (2006). The method of utilization of plant species was obtained from skilled and experienced older members of the tribal community.

## RESULTS AND DISCUSSION

A total of 39 plant species including herbs, shrubs, climbers and trees were frequently used by the Hajong tribe for curing ailments were recorded during the survey. The study revealed that herbs were dominant (49%) followed by trees (28%), shrubs (15%) and vines (8%) (Figure 2). Almost a similar trend was also observed by Ghani (2003) who conducted researches on other communities of Bangladesh and Halim et al. (2007) on the *Shiji* community in Southeastern Bangladesh but Mukul (2007); Miah and Chowdhury (2003) found that trees were dominant on a conservation area of Northern Bangladesh and Mro tribein Chittagong Hill Tracts, Bangladesh respectively.

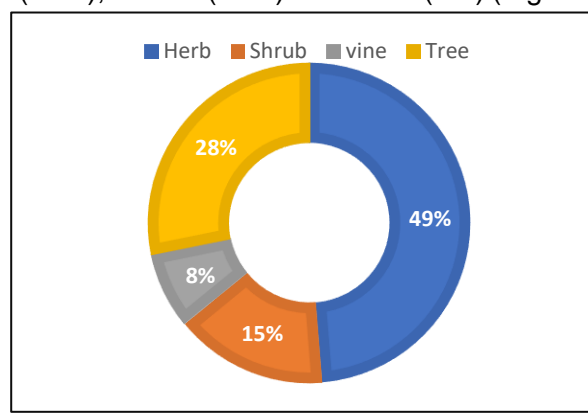

**Fig 2.** Percentage of the type of the medicinal plants identified from study area

Of the 39 plant species used by the TMPs, one could not be identified, suggesting that the hilly forested regions inhabited by the Hajongs may provide more plant species new to science. There were not too many ailments treated by the TMPs. The ailments treated included respiratory tract disorders, pain, cuts and wounds, urinary problems, jaundice, oral lesions, gastrointestinal disorders, tuberculosis, and itches. However, the Hajong healers were also cognizant of complicated diseases like cardiovascular disorders, cancer, and tumor and had specific medicinal plant formulations for their treatment. While cardiovascular disorders were recognized by the TMPs as patients having chest pain or irregular heartbeats, it could not be ascertained with certainty as to how the TMPs diagnosed cancer without any clinical diagnostic procedures. But they can't diagnose all of the symptoms without any doctor's help, in some cases they just refer it as just some illness, not any major disease, like; Cancer. However, in some instances cancer patients came from a nearby Christian missionary-run hospital where such patients were diagnosed properly. Tumors were defined by the TMPs as unexplained visible swellings; however, sometimes patients diagnosed with tumors were diagnosed as such at the nearby missionary run hospital. What was surprising was that even though the TMPs did not have any proper diagnostic procedures for the afore-mentioned diseases, they somehow had plant species and formulations, which they claimed to be effective treatments against these diseases.

**Table 1:** Medicinal plants and uses of The *Hajong Tribe*

| Sl. | Scientific Name | Family Name | Local Name | Used Part(s) | Ailment |
|---|---|---|---|---|---|
| 1 | *Justicia adhatoda* L. | Acanthaceae | Basok (G) | Leaf | Cough, Asthma, Phthisis, Malaria and Bleeding of piles |
| 2 | *Centella asiatica* | Apiaceae | Thankuni (G) | Leaf | Stomach ache |
| 3 | *Rauwolfia serpentina* | Apocynaceae | Unknown (G) | Root | Stomach ache |
| 4 | *Artemisia vulgaris* | Asteraceae | Naag Dana (G) | Leaf | Diarrhea. Dysentery |
| 5 | *Mikania micrantha* | Asteraceae | Unknown (G) | Leaf, whole plant | Flatulence |
| 6 | *Mikania cordata* | Asteraceae | Refugee lota (T) | Leaf | To stop bleeding from external cuts |
| 7 | *Alocasia macrorrhizos* | Araceae | Kochu (T) | Whole plant | Stomach Pain, Stomach disease |
| 8 | *Berberis thunbergii* | Berberidaceae | Khurmai (G) | Leaf | Blood loss |
| 9 | *Ananas comosus* | Bromeliaceae | Anaros (G) | Tender leaf | Stomach ache, Deworming |
| 10 | *Carica papaya* | Caricaceae | Pepe/ Pabda (G) | Leaf, Latex, Fruit | Dysentery, Peptic ulcer |
| 11 | *Terminalia chebula* | Combretaceae | Horitoki (W) | Fruit | Aversion |
| 12 | *Terminalia arjuna* | Combretaceae | Arjun (W) | Bark | Dysentery |
| 13 | *Terminalia bellirica* | Combretaceae | Bahera (W) | Fruit | Asthma, Cough, Dysentery |

| 14 | *Cuscuta reflexa* | Convolvulaceae | Sona lota (T) | Whole plant | Asthma, Flatulence |
|---|---|---|---|---|---|
| 15 | *Kalanchoe pinnata* | Crassulaceae | Pathor kuchi (G) | Whole plant | Cough, Flatulence |
| 16 | *Dillenia indica* | Dilleniaceae | Chalta/Choilta (T) | Fruit | Coughs, Fever |
| 17 | *Diospyros peregrina* | Ebenaceae | Gab (W) | Bark, Fruit | Dysentery, Cholera |
| 18 | *Swertia perennis* | Gentianaceae | Chirota (G) | Whole plant | Bad stomach |
| 19 | *Ocimum tenuiflorum* L. | Lamiaceae | Tulsi (G) | Leaf, stem | Coughs, Cancer, Tuberculosis |
| 20 | *Hyptis sauveolens* | Lamiaceae | Tokma (G) | Fruit | Flatulence, Acidity, Gastric troubles |
| 21 | *Cinnamomum verum* | Lauraceae | Daruchini (W) | Bark | Asthma, Coughs |
| 22 | *Punica granatum* | Lythraceae | Dalim/Dalum (G,T) | Fruit | Intestinal worms |
| 23 | *Hibiscus rosa-sinensis* | Malvaceae | Joba (G,T) | Leaf, Flower | Dysentery |
| 24 | *Azadirachta indica* | Meliaceae | Neem (G,T) | Leaf, stem | Skin disease |
| 25 | *Tinospora crispa* | Menispermaceae | Aamguruj (G,W) | Stem | Intestinal worms |
| 26 | *Streblus asper* | Moraceae | Sheora (W) | Leaf, Root | Stomach problems |
| 27 | *Syzygium aromaticum* | Myrtaceae | Long (W) | Flower bud | Asthma |
| 28 | *Dendrocalamus hamiltonii* | Poaceae | Tama baash (W) | Leaf, Stem | Stomach problems |
| 29 | *Phyllanthus emblica* | Phyllanthaceae | Amloki (W) | Fruit | Aversion, Bad stomach |
| 30 | *Ziziphus mauritiana* | Rhamnaceae | Boroi (T) | Leaf, Fruit | Dysentery, Diarrhea |
| 31 | *Aegle marmelos* | Rutaceae | Bel (T) | Fruit | Gastric, Flatulence |
| 32 | *Solanum torvum* | Solanaceae | Kalahota, Dhol baigon (G,W) | Fruit | Diarrhea. Dysentery |
| 33 | *Datura metel* | Solanaceae | Dhutura/Dhutra (G,W) | Leaf | Dysentery |
| 34 | *Clerodendrum infortunatum* | Verbenaceae | Bitu gach (W) | Leaf | Diarrhea |
| 35 | *Cissus quadrangularis* L. | Vitaceae | Harjora (G,W) | Stem | Pain due to bone fracture, Bone fracture |
| 36 | *Clerodendrum Vicosum* | Verbenaceae | Bhati/ Bhat (G,W) | Leaf, | Dysentery |
| 37 | *Zingiber officinale* | Zingiberaceae | Ada (G,W) | Rhizome | Asthma, Coughs |
| 38 | *Curcuma longa* | Zingiberaceae | Holud (G,W) | Leaf, Rhizome | Skin disease, Bad stomach |
| 39 | *Unknown* | Unknown | Somrit (G,W) | Fruit | Asthma, Cough |

*G: Grown in the garden; W: Wild; T: Topical

**Table 2:** Application and formulation of the plants

| Sl. | Scientific Name | Local Name | Ailment | Application and formulation |
|---|---|---|---|---|
| 1 | *Justicia adhatoda* L. | Basok | Cough, Asthma, Phthisis, Malaria and Bleeding of piles | The juice extracted from leaves bruised with water is drunk. |
| 2 | *Centella asiatica* | Thankuni | Stomach ache | Leaves of C. asiatica is ground well to prepare juice. The juice is taken together with goat's milk before breakfast |
| 3 | *Rauwolfia serpentina* | Unknown | Stomach ache | The roots are grounded and the decoction is fed. Two or three teaspoonfuls of leaf extract are taken directly |
| 4 | *Artemisia vulgaris* | Naag Dana | Diarrhea. Dysentery | Juice of the leaves is taken after breakfast or any food. |
| 5 | *Mikania micrantha* | Unknown | Flatulence | Squeezed leaves juice is taken 2-3 times a day after having food. |
| 6 | *Mikania cordata* | Refugee lota | To stop bleeding from external cuts | Mashed leaves is used direct in the affected area. |
| 7 | *Alocasia macrorrhizos* | Kochu | Stomach Pain, Stomach disease | The fleshy tuber is cut into small pieces and dried well. Thereafter, it is taken after cooking for some days. |
| 8 | *Berberis thunbergii* | Khurmai | Blood loss | Juice from the leaves is taken for 7-15 days once a day. |
| 9 | *Ananas comosus* | Anaros | Stomach ache, Deworming | Fruit juice is taken for fever. Young leaf is chewed for helminthiasis and jaundice |
| 10 | *Carica papaya* | Pepe/ Pabda | Dysentery, Peptic ulcer | Fruit is taken directly or juice for 2-3 days. |
| 11 | *Terminalia chebula* | Horitoki | Aversion | Fruits are soaked in water and then taken for constipation and vomiting. |
| 12 | *Terminalia arjuna* | Arjun | Dysentery | Soaked water of bark for the treatment of heart problem and burning sensations. Leaf juice to treat jaundice, dysentery. |
| 13 | *Terminalia bellirica* | Bahera | Asthma, Cough, Dysentery | Fruits are eaten for asthma and cough. |
| 14 | *Cuscuta reflexa* | Sona lota | Asthma, Flatulence | Whole plant Is formulated into juice and drank once a day. |
| 15 | *Kalanchoe pinnata* | Pathor kuchi | Cough, Flatulence | Juice of whole plant is taken for cure. |
| 16 | *Dillenia indica* | Chalta/ Choilta | Coughs, Fever | Fruit juice is taken for fever and coughs. |
| 17 | *Diospyros peregrina* | Gab | Dysentery, Cholera | Bark decoction is used for dysentery and cholera. |
| 18 | *Swertia perennis* | Chirota | Bad stomach | The whole plant is soaked over night and taken the water in the morning. |
| 19 | *Ocimum tenuiflorum* L. | Tulsi | Coughs, Cancer, Tuberculosis | Juice from macerated leaves to treat coughs [O]. Stems are worn as garland around the neck for tuberculosis. |
| 20 | *Hyptis sauveolens* | Tokma | Flatulence, Acidity, Gastric troubles | Fruits for the treatment of flatulence, acidity, gastric troubles. |
| 21 | *Cinnamomum verum* | Daruchini | Asthma, Coughs | Bark juice for asthma and coughs. |
| 22 | *Punica granatum* | Dalim/ Dalum | Intestinal worms | Ripe fruit and leaf juice for diabetis and intestinal worms. |
| 23 | *Hibiscus rosa-sinensis* | Joba | Dysentery | Whole plant extract for dysentery. |
| 24 | *Azadirachta indica* | Neem | Skin disease | Leaves are dried and powdered and taken every morning for allergy, eczema, skin diseases and diabetes. |
| 25 | *Tinospora crispa* | Aamguruj | Intestinal worms | Soaked water of the stem is used for worms. |
| 26 | *Streblus asper* | Sheora | Stomach problems | Bark juice is taken for medication. |
| 27 | *Syzygium aromaticum* | Long | Asthma | Flower bud to treat asthma and coughs. |
| 28 | *Dendrocalamus hamiltonii* | Tama baash | Stomach problems | Tender plant is used for stomach problems. |
| 29 | *Phyllanthus emblica* | Amloki | Aversion, Bad stomach | Fruit is taken for stomach problems 2-3 times a day. |
| 30 | *Ziziphus mauritiana* | Boroi | Dysentery, Diarrhea | Fruit and leaf is taken everyday for cure. |
| 31 | *Aegle marmelos* | Bel | Gastric, Flatulence | Fruit and leaf juice is used for gastric problems. |
| 32 | *Solanum torvum* | Kalahota, Dhol baigon | Diarrhea. Dysentery | Whole plant is take for diarrhea. |
| 33 | *Datura metel* | Dhutura/ Dhutra | Dysentery | Flower and seed for cold and nervous disorders. Crushed leaf is applied to painful areas. |
| 34 | *Clerodendrum infortunatum* | Bitu gach | Diarrhea | Diarrhea. Juice obtained from macerated leaf is taken once daily for 2 days. |

| 35 | *Cissus quadrangularis* L. | Harjora | Pain due to bone fracture, Bone fracture | Paste of stem or rhizome is mixed with grounded ginger and applied as a poultice to the fractured area. |
| --- | --- | --- | --- | --- |
| 36 | *Clerodendrum Vicosum* | Bhati/ Bhat | Dysentery | Juice from whole plant is taken 2-3 times a day. |
| 37 | *Zingiber officinale* | Ada | Asthma, Coughs | Juice from tuber or root is taken several times till it cures. |
| 38 | *Curcuma longa* | Holud | Skin disease, Bad stomach | Root is smashed and applied in the infected area. |
| 39 | *Unknown* | Somrit | Asthma, Cough | Whole plant is used. |

In the **Table 1** plants are listed in order to their family names. This table shows only the plants blogical informations. In **Table 2** all the formulations are described as the hajong people dscribed it.

It was evident from the study that Hajong tribe was greatly dependent on herbal medicines. In now a days they are leaning forward to the mordern alopathical medicines but a large number of tribal people are still counting on the TMPs. They are still a believer of those medicines. TMPs are always in the urge of helping their people but in some cases they can not always help. They reffer to the general hospital but very few goes their. As they are always dependent on medicinal plants and Kabiraj or Tribal medicine practitioners, they get actual medical help avery little. But in terms of first aids and general help, they get it.

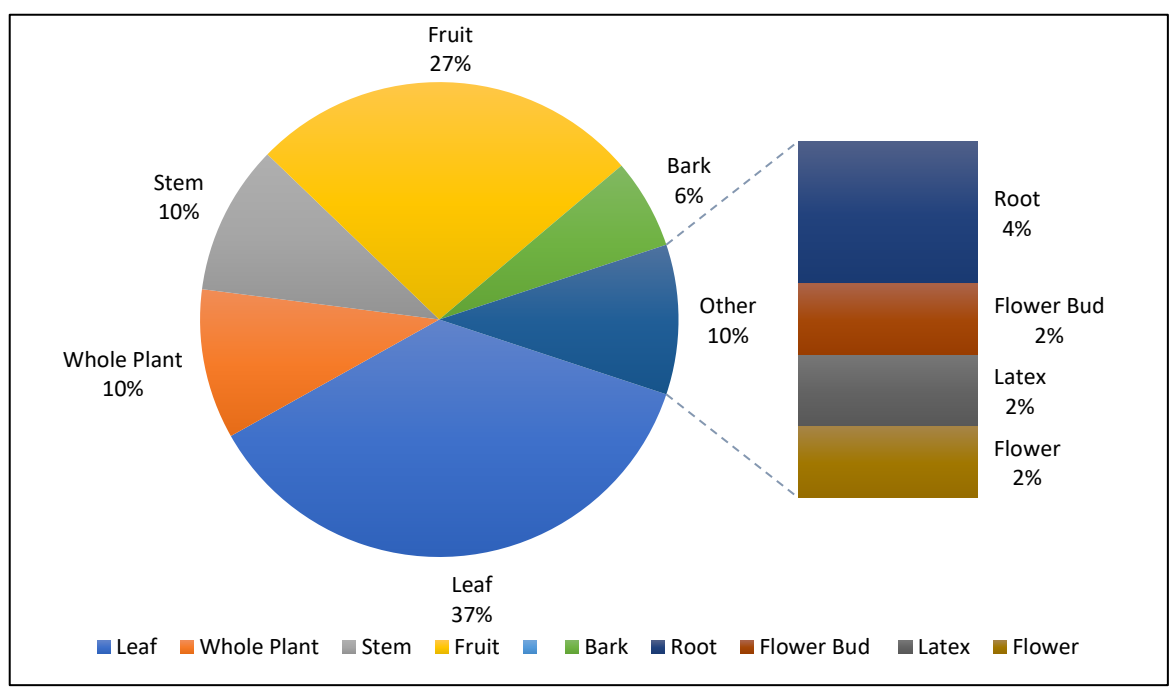

**Fig 3.** Medicinal plants break up by used parts

In our search I have found that, they use 39 plants for 12 types of diseases. Mainly for stomach problems. Of the 39 plants only, a few were applied on the affected area, most of them were taken orally. They take this medicine either by themselves or made by the TMPs. But for curing the trust the source.

In most of the houses I found a small garden consist of some herbal plants they use the most. *Artemisia vulgaris, Rauwolfia serpentina, Centella asiatica, Justicia adhatoda, Kalanchoe pinnata, Swertia perennis, Ocimum tenuiflorum* were very common in their house.

*Ocimum tenuiflorum* is largely used and is also a religious plant in the Hindu houses. It helps in various diseases that the tend to cultivate it on their own. *Justicia adhatoda* is also a very common plant in the area. As it grows on its own in the pristine lands of the village or near the place they live, they collect the plat from here and there. But in some houses, they grow it of their own.

In all the listed plants very, few is cultivated (**Fig 4**). Only the herbs are cultivated or collected from local areas. But the trees are very rare to find and a small number of people knows the actual use of the plant. For this they go to near jungle or where the tree is found, to collect the used parts to make medicine.

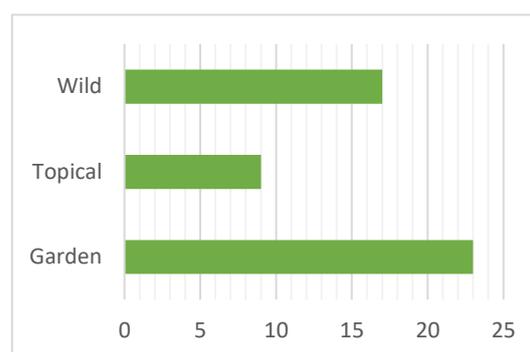

**Fig 4**. Plants by their habitat

The use of *Adhatoda vasica* for treatment of piles bleeding found in the study area. Same result also found by Uddin et al. (2008), Mukul et al. (2007), Miah and Chowdhury (2003) in Bangladesh and Sajem and Gosai (2006) also found same result in Jaintia tribes in Northeast India. The use of Centella asiatica against piles has been also reported in the study area while the same species is used against dysentery (Miah and Chowdhury, 2003), dysentery, diarrhea and gastric pain (Mukul et al., 2007) in Bangladesh but Sajem and Gosai (2006) reported against conjunctives, eye injury, stomachic, indigestion and flatulence in Jaintia tribes in Northeast India.

## CONCLUSION

In the study I found a large number of medicinal plants used by the tribal people. The Pi chart shows which part of the plant they use the most. I have found that, the leaves were the most. I need to investigate more to find the medicinal value of the plants. This could open an enormous gate of medicinal data that could help mankind. The dependence of health care practice through locally available practitioners of the Hajong, has shown a particular pattern of forest resources exploitation and an extreme dependence on forests. The information generated from the present study regarding the medicinal plant use by the Hajong tribes need a thorough phytochemical investigation including alkaloid extraction and isolation along with few clinical trials. This could help in creating mass awareness regarding the need for conservation of such plants and also in the promotion of ethno-medico- botany knowledge within the region besides contributing to the preservation and enrichment of the gene bank of such economically important species before they are lost forever.

The information generated from the present study regarding the medico-religious plants used by the Hajong tribes need a thorough phytochemical investigation. This could help in creating awareness regarding the need for conservation of such plants and also in the promotion of ethno-medico-botany knowledge within the region besides contributing to the preservation and enrichment of the gene bank of such economically important species before they are lost forever. Traditional culture in different Hajong populated areas is very fast declining with lot of traditional knowledge under the influence of dominant culture. Cultural diversity conservation is needed urgently.